\documentclass[aps,physrev, twocolumn,superscriptaddress]{revtex4-2}

\usepackage[dvipsnames]{xcolor}
\usepackage{lipsum}
\usepackage{amsmath}
\usepackage[colorlinks=true, allcolors=blue]{hyperref}
\usepackage{graphicx}
\usepackage[english]{babel}
\usepackage{dcolumn}
\usepackage{bm}
\usepackage{amssymb}
\usepackage{float}
\usepackage[T1]{fontenc}
\usepackage{physics}
\usepackage{comment}

\begin{document}

\preprint{APS/123-QED}

\title{Activity driven buckling and pattern formation in  shells of oriented solids}

\author{Niels de Graaf Sousa}
\thanks{These authors contributed equally to this manuscript.}
\author{Varun Venkatesh}
\thanks{These authors contributed equally to this manuscript.}
\author{Amin Doostmohammadi}
\email[]{doostmohammadi@nbi.ku.dk}
\affiliation{Niels Bohr Institute, University of Copenhagen, 2100 Blegdamsvej 17, Copenhagen, Denmark}

\date{\today}

\begin{abstract}
    We investigate shells of active oriented solid, materials in which orientationally ordered active particles are embedded in a deformable elastic surface. Focusing on cylindrical geometries, we show that active stresses drive a new class of buckling instabilities and nonlinear patterns absent in passive shells. Linear stability analysis reveals that the unstable buckling mode is selected by the nematic orientation and activity sign, leading to axial, circumferential, and helical deformations. Remarkably, circumferential modes become unstable at arbitrarily small activity due to the absence of stretching costs. The results of the linear stability analysis are corroborated by full nonlinear simulations, which further uncover steady diamond shaped patterns and persistent dynamical states including oscillations, traveling domain walls, and propagating waves. Our results establish fundamental buckling modes and emergent patterns in shells of active oriented solid materials, with potential relevance to active biological tissues and engineered responsive materials.
\end{abstract}

\maketitle

Mechanical instabilities are a universal route by which materials generate structure and function across scales, from wrinkling membranes and growing tissues to shape morphing metamaterials \cite{kim_hierarchical_2011, huang_differential_2018, ben_amar_wrinkles_2025, dudek_shape-morphing_2025}. In passive systems, the interplay between geometry and elasticity gives rise to a rich spectrum of buckling transitions and pattern formation \cite{holmes_elasticity_2019, wang_mechanics_2022}. Active materials, whose constituents continuously generate internal stresses, introduce an additional route to shape formation that remains far less understood, particularly in solids where activity acts within an elastic network rather than a flowing fluid \cite{volvox_embryo,elephant_trunk,guillamat2026guidance}.

A particularly important class of active materials consists of systems with orientational order, where anisotropic constituents generate direction dependent stresses and mechanical responses \cite{AOS_Ramaswamy,AOS_Marchetti}. While the coupling between orientational order and the shape changes of materials is well recognized~\cite{maroudas-sacks_topological_2021, coen_mechanics_2023, xiong_interplay_2014}, theoretical studies have overwhelmingly focused on fluids, first by exploring dynamics on rigid or prescribed geometries~\cite{flows_lsa_cylinder,bell2022active,pearce2020defect}, and more recently on dynamical deformable surfaces~\cite{Berthoumieux_2014,metselaar2019topology,thijssen2021activity,Al-Izzi_Morris_2023,nejad2025thin,mietke_morpho,venkatesh_emergent_2026}. Yet many biological and synthetic materials instead behave as active solids, including epithelial tissues~\cite{lampart2025morphometry,benoit_epithelia_solid}, endothelial tubes~\cite{endothelial_tubes}, cytoskeletal structures~\cite{chen_ricci}, muscle fibers~\cite{muscle_shankar}, and engineered responsive materials~\cite{soft_robotics,kotikian20183d}, where orientational order and internally generated stresses are directly coupled to shape changes and mechanical deformation. Despite this broad relevance, the mechanics of deformable active oriented solid shells remains essentially unexplored.

Here, we study instabilities and deformation modes of active nematic solids confined to deformable cylindrical shells and show that activity fundamentally governs shell stability and pattern selection. We demonstrate a new class of activity-driven buckling instabilities in which the unstable mode is selected by the nematic orientation and the sign of activity, giving rise to axial, circumferential, and helical deformations. Remarkably, some modes become unstable at arbitrarily small activity due to the absence of stretching costs. Beyond the linear instability, nonlinear simulations reveal self organized diamond patterns and persistent dynamical states including oscillations and traveling waves generated through feedback between strain and nematic alignment.\\

\noindent\textbf{Model.--}We consider a cylindrical shell populated by oriented active particles that are capable of exerting stress and realigning based on the deformation strain of the surface. The dynamics of the thin shell nematic with order parameter $Q^{ij}$ and radius $R$ are governed by the free energy functional $\mathcal{F} = \int (f_{\text{shell}} + f_{\text{nematic}} + f_{\text{coupling}}) \, dS$:

\begin{equation}
   \begin{aligned}
\label{eq:free-energy}
&\mathcal{F} =\frac{1}{2} \int dS\bigg[ D (\nabla^2 w)^2 + \frac{Eh}{(1+\nu)} \left( \epsilon^{ij}\epsilon_{ij} + \frac{\nu}{1-\nu} (\epsilon^k_k)^2 \right)   \\
& -A(1 - \frac{1}{2}{}\operatorname{Tr}(Q^2))\operatorname{Tr}(Q^2) +L(\nabla_k Q^{ij})^2 + 2\lambda Q^{ij} \epsilon_{ij} \bigg] ,
\end{aligned} 
\end{equation}
where $\epsilon_{ij} = \frac{1}{2}(\nabla_i u_j + \nabla_j u_i) - b_{ij} w + \frac{1}{2}(\nabla_i w)(\nabla_j w)$ is the nonlinear strain tensor, $b_{ij}$ is the curvature tensor, $L$ is the Frank elastic constant, $A$ is the nematic ordering strength, and $D = \frac{Eh^3}{12(1-\nu^2)}$ is the flexural rigidity, which encapsulates the material properties, $E$ is the Young's modulus, $\nu$ the Poisson's ratio, and $h$ denotes the shell thickness. The strength of the nematic-shell coupling is set by $\lambda$. In the overdamped limit, the in-plane $u^j \in \{u,v\}$ and radial $w$ displacements read:
 \begin{align}
    \label{eq: non-linear governing equations}
    \eta \partial_t u^j &= h \nabla_i \sigma^{ij}_{\text{tot}}, \\
    \label{eq: non-linear governign eqaution radial}
    \gamma \partial_t w &= -D \nabla^4 w + h \nabla_j (\sigma^{ij}_{\text{tot}} \nabla_i w) + h \sigma^{ij}_{\text{tot}} b_{ij} + p.
\end{align}

Here $p$ is the external radial pressure and the effective total stress is defined as $\sigma^{ij}_{\text{tot}} = \sigma^{ij}_{\text{el}} - \tilde{\zeta} Q^{ij}$, where the elastic stress is derived from the free-energy functional as $\sigma^{ij}_{\text{el}} = \delta \mathcal{F} / \delta \epsilon_{ij}$ and $-\tilde{\zeta} Q^{ij}$ describes the active stress contribution. Since the total stress contains two terms linear in the nematic order parameter, we define an effective activity $\zeta =\tilde{\zeta} - \lambda$ to capture the total deviatoric stress exerted by the nematic on the solid. The sign of $\zeta$ distinguishes extensile ($\zeta > 0$) from contractile ($\zeta < 0$) active systems~\cite{hatwalne2004rheology,marchetti2013hydrodynamics}.  Assuming no convection or co-rotation, the nematic tensor evolves in the reference frame of the deforming surface, driven by the total molecular field $H^{ij}$:
\begin{equation}
    \partial_t Q^{ij} =  \Gamma  H^{ij},
\end{equation}
where $\Gamma$ is the rotational diffusivity. Using $g_{ij}$ as the metric tensor of the surface, the molecular field for curved surfaces is $H^{ij} = -\frac{1}{\sqrt{g}}g^{ik}g^{j \ell }\frac{\delta \mathcal{F}}{\delta Q^{kl}}$.\\
\begin{figure*}[!t]
    \centering
    \includegraphics[width=\linewidth]{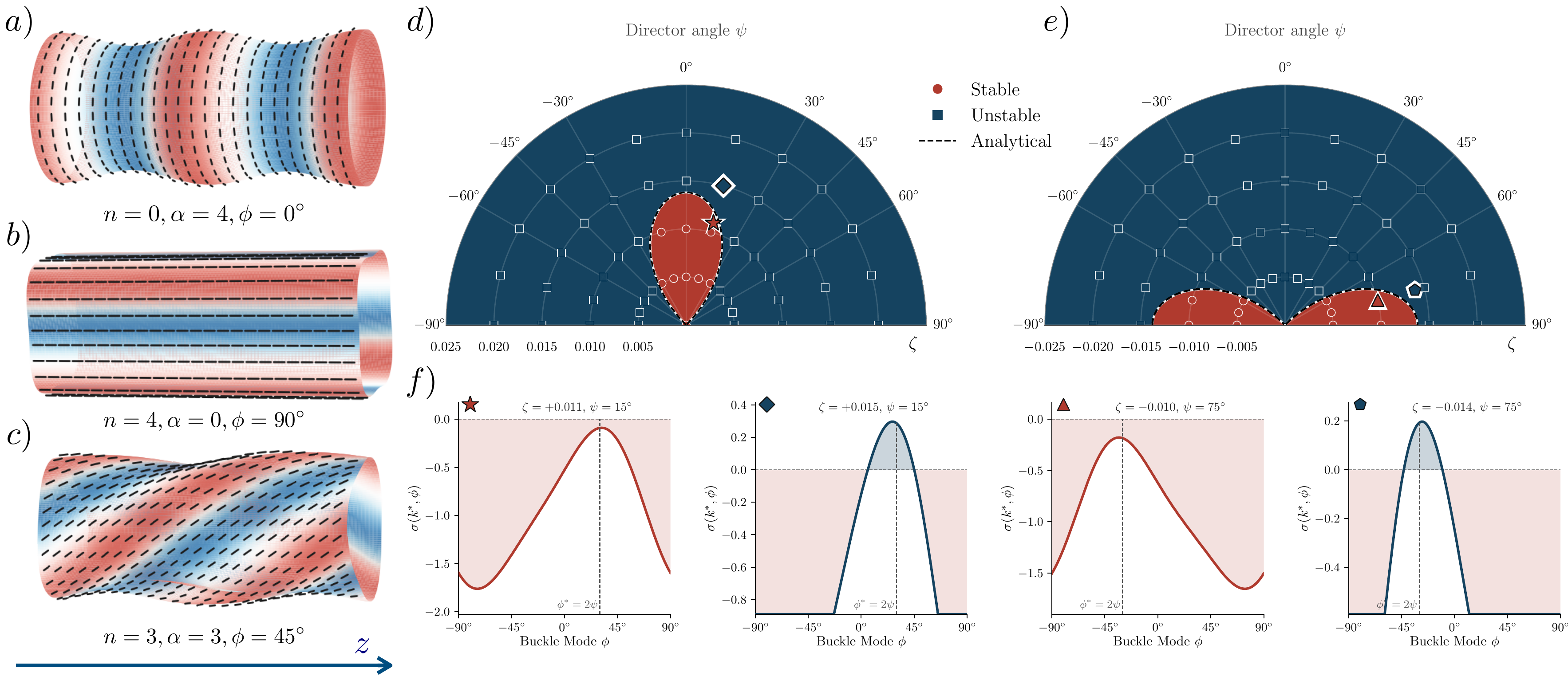}
    \caption{\textbf{ 
    Activity-induced buckling instability of a cylindrical shell of oriented solid.}
    \textbf{(a)} Axial mode ($n=0$, $\alpha=4$): purely $z$-dependent deformation forming axial stripes, with director aligned along the azimuthal direction ($\psi = \pm 90^\circ$).
    \textbf{(b)} Circumferential mode ($n=4$, $\alpha=0$): purely $\theta$-dependent deformation forming circumferential rings, with director along the cylinder axis ($\psi = 0^\circ$).
    \textbf{(c)} Helical mode ($n=\alpha=3$): combined deformation forming helices at $\phi=45^\circ$, with director along the opposite  direction ($\psi = -45^\circ$).
    Here $n$ is the azimuthal wavenumber, $\alpha$ the axial wavenumber, and $\phi = \arctan\!\big[n/\alpha\big]$ the angle of the buckle wavevector with the cylinder axis.
    Black directors represent the nematic director on the deformed surface of the contractile cylindrical shells.
    \textbf{(d)} Polar phase diagrams for extensile ($\zeta>0$) and \textbf{(e)} contractile ($\zeta<0$) activity.
    Filled regions indicate the numerical prediction of stable (red) and unstable (navy) states; the black dashed curve is the analytical critical activity (Eq. \ref{eq: critical activity final }).
    Markers denote non-linear simulation outcomes, circles (stable) and squares (unstable), showing close agreement with the analytical boundary.
    \textbf{(f)} Growth-rate spectra $\sigma(k^*,\phi)$ at the four labeled $(\zeta,\psi)$ points, evaluated at the critical wavenumber $k^*$. We set $h=1.2 \times 10^{-5}$.
    }
    \label{fig: analytics}
\end{figure*}

\noindent\textbf{Buckling instability.--}To gain analytical insight into the mechanism by which active stress destabilizes the cylindrical shell, we perform a linear stability analysis of the reference configuration, taking the director to be uniformly time-independent orientation at an angle $\psi$ with respect to the $z$-axis, which satisfies the nematic symmetry $\psi=\psi +\pi$. The prebuckling state is determined by the applied loads $N_{zz}^0$, $N_{z\theta}^0$, and $N_{\theta\theta}^0$, expressed via the {\it Airy} stress function $\Phi$ defined as $N_{zz}= \frac{1}{R^2}\frac{d^2 \Phi}{d \theta^2},\, N_{\theta \theta}= \frac{d^2 \Phi} {d z^2}\, \text{and} \,N_{z \theta}=-\frac{1}{R}\frac{d^2 \Phi}{d z d \theta}$. For vanishing initial displacements, $(w_0, u_0, v_0) = 0$, these reduce to the active stress resultants $N^0_{ij} = h \sigma^a_{ij}$. To ensure radial force balance, a radial pressure $p$ is applied, which is given by $N_{\theta \theta}^0=-pR$. Under this condition, the cylinder is in a steady state and linearizing Eqs.~\ref{eq: non-linear governing equations} and \ref{eq: non-linear governign eqaution radial} leads to the DMV theory for thin cylindrical shells \cite{Yamaki1984,Donnell1933} (see End Matter for details). To assess the stability of the prebuckled state, we introduce perturbations to the radial displacement $w$ and the applied loads $\Phi$,
\begin{align}
    w= \delta w \quad \delta w= A \exp{[i( \alpha z+ n \theta)+\sigma t]}\\
    \Phi=\Phi^0+\delta \Phi\quad \delta \Phi= B \exp{[i( \alpha z+ n \theta)+\sigma t]}.
\end{align}
where $\alpha$ is the axial wavenumber and $n\in\mathbb{N}$ the azimuthal mode number. We define the combined wavenumber as $k^2 = \alpha^2 + \frac{n^2}{R^2}$. Applying the perturbation to the governing equations (Eq.~\ref{eq: non-linear governing equations} and \ref{eq: non-linear governign eqaution radial}) gives,
\begin{align}
    \label{eq: pertubed state}
    \nabla^4 \delta \Phi+\frac{Eh}{R}\frac{d^2 \delta w}{dz^2}=0 \\
    \begin{aligned}
        \label{eq: perturbed radial displacment}
        \gamma \frac{d \delta w}{dt} + D\nabla^4 \delta w - \frac{1}{R}\frac{d^2 \delta \Phi}{d z^2}
  = N_{zz}^0\frac{d^2 \delta w}{d z^2}\\
   + \frac{2N_{z\theta}^0}{R}\frac{d^2 \delta w}{d z\,d\theta}
   + \frac{N_{\theta\theta}^0}{R^2}\frac{d^2 \delta w}{d\theta^2}.
    \end{aligned}
\end{align}

The compatibility equation (Eq.~\ref{eq: pertubed state}) leads to the following relation between the perturbation amplitudes $B=\frac{Eh\alpha^2}{Rk^4}A$. Inserting this relation into the radial displacement equation (Eq.~\ref{eq: perturbed radial displacment}) leads to the following growth rate, 
\begin{equation}
    \label{eq: growth rate non dimensionalized}
    \gamma \sigma = -Dk^4 - \frac{Eh\alpha^4}{R^2 k^4}
  - N_{zz}^0\alpha^2
  - \frac{2N_{z\theta}^0\,\alpha n}{R}
  - \frac{N_{\theta \theta}^0\, n^2}{R^2}.
\end{equation}

Here we introduce the following dimensionless variables to express the growth rate in dimensionless form $\zeta'=\frac{\zeta}{E}$, $h'=\frac{1}{12(1-\nu^2)}\left(\frac{h}{R}\right)^2$, $\alpha'=\alpha R$ ,  $\sigma'=\frac{\sigma \gamma R^2}{h E}$ and $n'=n$. Dropping the primes, introducing the active stress and the buckle propagation angle $\phi = \arctan(n/\alpha)$ the growth rate in dimensionless form reads

\begin{equation}
    \sigma = -h k^4 - \cos^4\phi
             + \frac{\zeta k^2}{2}\cos\!\big(2[\psi-\phi]\big).
    \label{eq:growthrate_phi}
\end{equation}

When the active term in Eq.~\ref{eq:growthrate_phi} dominates over curvature-induced elasticity contributions, the buckle-mode angle that maximizes the growth rate is $\phi^{*}=\psi$ for extensile activity and $\phi^{*}=\psi\pm\pi/2$ for contractile activity, consistent with the flat-plate result (see End Matter). In both cases the buckle wave vector aligns with the principal axis of the active stress along which the shell is most compressed, recovering the standard buckling intuition in the active setting: for contractile, axis-perpendicular directors select axial modes, axis-parallel aligned directors select circumferential modes, and directors at intermediate angles select helical modes (see schematic in Fig.~\ref{fig: analytics}a,b and c). Having shown how the systems buckles when activity dominates, we now turn to the structure of the buckling mode selected at onset of the instability, where the growth rate vanishes,

\begin{equation}
    \label{eq: crticial nondim activity}
    \zeta_{c}=\frac{2}{k^2}\frac{hk^4+\cos^4\phi}{\cos\!\big(2[\psi-\phi]\big)}.
\end{equation}

The numerator of Eq.~\ref{eq: crticial nondim activity} is always positive, so the sign of the denominator determines which modes become unstable in the extensile and contractile cases for each orientation angle $\psi$. Buckling is driven by extensile activity for modes within $\phi = \psi \pm \pi/4$ around the nematic director, and by contractile activity for modes rotated by $\pi/2$, i.e.\ $\phi = \psi + \pi/2 \pm \pi/4$. The two cases are related by the symmetry $(\zeta, \psi) \rightarrow (-\zeta, \psi + \pi/2)$. To validate the analytical predictions, we carried out non-linear simulations across a range of activities and director angles (see End Matter for the details of the simulations). The simulation outcomes (blue squares: unstable; red circles: stable, Fig.~\ref{fig: analytics}d,e) align well with the analytically predicted stability boundary.

The growth rate exhibits a non-trivial dependence on the perturbation eigenmodes and eigenvalues (Eq.~\ref{eq:growthrate_phi}). As such, to gain further analytical insight, we consider a number of limiting cases. First, we consider the case where the second stabilizing term $\cos^4\phi$ vanishes for any purely circumferential mode $\phi = \pm\tfrac{\pi}{2}$, since these modes have no axial variation and therefore incur no membrane-curvature energy.  In this case the growth rate reduces to

\begin{equation}
    \sigma\big|_{\phi=\pm\frac{\pi}{2}}
    = -h k^4 -\frac{\zeta k^2}{2} \cos{(2 \psi)}.
    \label{eq:growthrate_circ}
\end{equation}

For long-wavelength perturbations the active term scales as $k^2$ while the stabilizing bending term scales as $k^4$, so the active term always destabilizes at sufficiently small $k$.  Consequently, the cylinder will buckle circumferentialy at arbitrarily small activity whenever the active contribution in Eq.~\ref{eq:growthrate_circ} is positive. This occurs for \ $|\psi| > \tfrac{\pi}{4}$, for extensile systems, and for \ $|\psi| < \tfrac{\pi}{4}$, for contractile systems.

Secondly, we find that the critical wave-vector that minimizes the activity to rise the instability is given by $k^*{}^2= \cos^2{\phi}/\sqrt{h}$. Incorporating the critical wave-vector and minimizing with respect to the buckle mode we obtain that the activity is minimized when $\phi^*=2 \psi$ which incorporating into Eq.~\ref{eq: crticial nondim activity} yields the following critical activity,

\begin{equation}
    \label{eq: critical activity final }
    \zeta_{c}=4 \sqrt{h} \cos{(2 \psi)}.
\end{equation}

The critical activity thus traces a set of rhodonea or rose curves in the $(\zeta,\psi)$ plane (\textit{dashed black line} in Fig.~\ref{fig: analytics}d,e), in agreement with the lobed structure recovered from the numerical solution of the growth rate (\textit{white line} in Fig.~\ref{fig: analytics}d,e). We further plot the growth rate in two distinct regions near the instability threshold, confirming that the mode driving the instability is indeed $\phi^{*} = 2\psi$ (Fig.~\ref{fig: analytics}f).\\

\begin{figure*}[!t]
    \centering
    \includegraphics[width=0.9\linewidth]{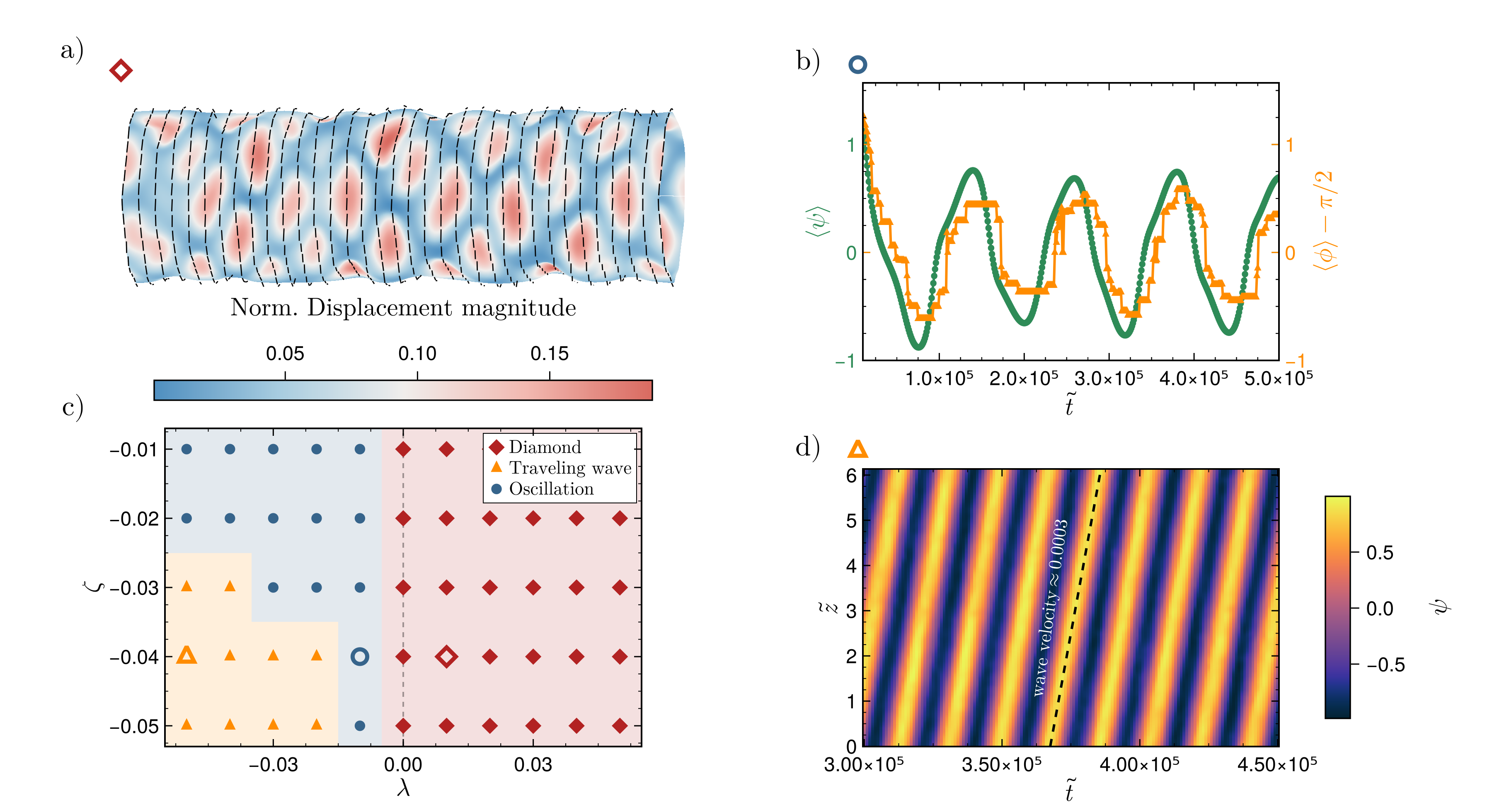}
    \caption{\textbf{Dynamic patterns on deformable shells of active nematic solids.} \textbf{(a)} Simulated snapshot of the nematic director surface deformation corresponding to a diamond pattern for  parameters $\zeta=-0.04$ and $\lambda=0.01$. \textbf{(b)} Time series of the coupled oscillatory dynamics for the nematic (green circles) and deformation (orange triangles) orientations for parameters $\zeta, \lambda = -0.04,-0.01$. \textbf{(c)} Phase diagram showing structural states of oscillatory stripes, traveling waves, and stationary diamonds in the $(\lambda, \zeta)$ parameter space. \textbf{(d)} Kymograph of the nematic orientation along a 1D slice, showing a traveling wave moving along the cylinder axis ($\zeta, \lambda = -0.04,-0.05$). We use dimensionless quantities of $\tilde{z}=z/R$, $\tilde{t} = tA\Gamma$ and normalize $\zeta$ and $\lambda$ by the Young's modulus $E$. Animations of the simulations are found in the \href{https://drive.google.com/drive/folders/1ob7kND11Z73RcO3rc9187QMQdbNS7JOB?usp=sharing}{supplementary movie}.
    }
    \label{fig:dynamics_states}
\end{figure*}

\noindent\textbf{Pattern selection.--}Beyond confirming the onset of instability, further long-time non-linear simulations show the emergence of two-mode diamond patterns (Fig.~\ref{fig:dynamics_states}a) with sufficient activity. These diamond patterns are found to be a common occurrence with the introduction of two-way coupling between the solid's deformation and the nematic director. The coupling parameter $\lambda$, when positive, causes the nematic to align with the local strain field. When the active nematic deforms the solid, the resulting strain simultaneously reorients the nematic. Without this coupling ($\lambda = 0$), the system is restricted to static diamond patterns; however, the feedback allows the system to sustain persistent, dynamic, out-of-equilibrium states.

The phase boundaries between these states are controlled by the coupling strength $\lambda$ and activity $\zeta$ (Fig.~\ref{fig:dynamics_states}c). For $\lambda \ge 0$, the feedback is stabilizing, and the system locks into stationary diamond patterns.  As seen in the previous section, sufficient contractile activity always drives buckle modes perpendicular to the director, with deformations aligned along it. Because $\lambda > 0$ aligns the nematic along this same direction, it sets a single stable axis. The nematic director thus ceases to change on a time scale relevant for deformations, and the dynamics reduces to those of a system driven by a slowly varying stress field, as is the case for $\lambda=0$.

Conversely, for $\lambda < 0$, the coupling introduces frustration between the preferred direction of active stress-induced deformation and the preferred nematic alignment. The director prefers to align perpendicular to the deformation and thus reorients, forcing the deformation field to also reorient towards this new direction. This continuous feedback destabilizes the steady-state configuration and drives the system into dynamic regimes, ultimately giving rise to a spatiotemporal phase of bulk oscillations, where surface deformation and nematic orientations switch periodically and uniformly, with a small time lag between them (Fig.~\ref{fig:dynamics_states}b). 

Increasing the magnitude of the activity $|\zeta|$ and coupling strength $|\lambda|$ results in the emergence of traveling waves in both the nematic and displacement fields, which is most readily observed in the orientation of the nematic order parameter, as seen in the kymograph in Fig.~\ref{fig:dynamics_states}d. This transition is driven by two factors. First, increasing the active stress reduces the characteristic length scale of the deformations. Second, strong coupling significantly shortens the time scale for nematic reorientation. Consequently, we observe the formation of heterogeneous domains that localize into bend and splay walls prior to the onset of larger domains in the deformation with perpendicular, almost zigzag-like structures. Once established, this pattern becomes dynamic due to the underlying frustration of the coupling. The persistent phase lag between the nematic rotation and deformation prevents an equilibrium and results in propelling the entire coupled structure forward as a traveling wave (See SI Movie \cite{SIMovies}).\\

\noindent\textbf{Discussion.--}
%Our results establish that active nematic stresses fundamentally govern the stability and pattern selection of cylindrical elastic shells. In the linear regime, the buckling mode is not set by geometry alone but is encoded in the nematic orientation and the sign of activity: extensile and contractile systems select orthogonal instability axes, with the critical activity tracing rhodonea curves in parameter space. Outside this regime, circumferential modes admit instabilities at arbitrarily small activity, since such deformations incur no bending energy penalty from the membrane. Beyond onset, nonlinear dynamics driven by strain-nematic feedback produce a rich variety of out-of-equilibrium states: static diamond patterns, bulk oscillations, and spatially propagating waves, none of which are accessible in passive shells or in active systems without two-way mechanical coupling.\\
These findings offer a new perspective on the mechanics of active cells confined to tubular geometries. Many biological processes involve oriented, mechanically active cells lining tubular structures, from epithelial and endothelial tubes~\cite{shyer_villification_2013,Tube_morphogenesis,endothelial_tubes} to cytoskeletal assemblies~\cite{gorfinkiel2011dynamics}, where the interplay between cellular activity, orientational order, and surface curvature governs collective behavior. Our results suggest that purely mechanical feedback between active stress generation and structural reorientation can drive a rich variety of spatiotemporal dynamics in such systems, without invoking biochemical pathways. Peristalsis, the coordinated propagation of contractile waves through tubular organs such as the intestine~\cite{muller2014crosstalk,spencer2020enteric} and reproductive tract~\cite{maia1970peristalsis,ezzati2014tubal}, is one prominent example where such mechanical self-organization may play a contributing role alongside biochemical signaling. More broadly, the activity-orientation-geometry interplay identified here may underlie a wider class of tubular tissue dynamics, a hypothesis that could be tested by selectively perturbing mechanical and biochemical pathways in experimental systems.

Beyond biology, our framework has direct implications for the design of engineered active materials. Responsive shells and soft actuators increasingly exploit anisotropic internal stresses to achieve programmable shape changes~\cite{Yao_2024,nematic_design_shape_morphing}. The present work provides a systematic map, encoded in the rhodonea stability diagram, of which buckling modes are accessible for a given material anisotropy and activity level. Crucially, the ability to select axial, circumferential, or helical deformations by tuning the nematic orientation offers a design principle that is both geometric and material-independent, applicable to any system in which oriented active elements are embedded in a deformable elastic surface.

%More broadly, our results highlight active oriented solid shells as a distinct and largely unexplored class of active matter, sitting between active fluids on rigid surfaces and passive elastic shells. Unlike active fluids, where activity reorganizes orientation without substantially deforming the substrate, active solids couple stress generation directly to shape change. Unlike passive shells, the material itself supplies the driving force for instability. 
Extending the present approach to non-cylindrical geometries such as the heart~\cite{3-d_printed_ventricles,Jin_Jie2025}, bladder~\cite{lampart2025morphometry}, or developing brain \cite{Chavoshnejad_2026,10.7554/eLife.107138}, where curvature gradients and topological constraints will further enrich the instability landscape, promises to uncover a new class of non-equilibrium morphodynamic phenomena at the intersection of active matter physics and tissue mechanics.

\begin{acknowledgments}
We thank Farzan Vafa for helpful discussions.  A. D. acknowledges funding from the Novo Nordisk Foundation (grant No. NNF18SA0035142 and NERD grant No. NNF21OC0068687), Villum Fonden (Grant no. 29476), and the European Union (ERC, PhysCoMeT, 101041418). Views and opinions expressed are however those of the authors only and do not necessarily reflect those of the European Union or the European Research Council. Neither the European Union nor the granting authority can be held responsible for them. The Tycho supercomputer hosted at the SCIENCE HPC center at the University of Copenhagen was used for supporting this work.
\end{acknowledgments}

% \section{Appendix}

\bibliography{apssamp}

\section*{End Matter}
\appendix

%\section{Isothermal Coordinates}

\section{Bridging the nonlinear equations and the linear analytical theory}

For completeness, we show in this appendix how the nonlinear equations used in the numerical simulations reduce, in the appropriate limit, to the linearized description underlying our analytical stability analysis. Radial force balance in simulations is given by Eq. \ref{eq: non-linear governign eqaution radial}. This equation reduces to the DMV form under three simplifications. First using the Leibniz rule we split the divergence,
\begin{equation}
    \nabla_j\!\left(\sigma^{ij}_{\text{tot}}\nabla_i w\right)
    = \left(\nabla_j\sigma^{ij}_{\text{tot}}\right)\nabla_i w
    + \sigma^{ij}_{\text{tot}}\,\nabla_j\nabla_i w.
\end{equation}
The first term drops because at linear order the pre-buckling stress is uniform $\nabla_j\sigma^{ij}_{\text{tot}}=0$. On the reference cylinder the only nonzero physical component of the second fundamental form is $b_{\theta\theta}=1/R$, hence
\begin{equation}
    h\,\sigma^{ij}_{\text{tot}}\,b_{ij} = \frac{N_{\theta\theta}}{R},
    \qquad N^{ij}\equiv h\sigma^{ij}_{\text{tot}}.
\end{equation}

Finally  expanding the covariant Hessian on the undeformed cylinder gives, $\nabla_z\nabla_z w = \partial^2_z w,\,\nabla_z\nabla_\theta w = \frac{1}{R}\partial_z \partial_\theta w,\,\nabla_\theta\nabla_\theta w = \frac{1}{R^2}\partial_\theta^2w,$
so that
\begin{equation}
    h\,\sigma^{ij}_{\text{tot}}\,\nabla_j\nabla_i w
    = N_{zz}\frac{\partial^2 w}{\partial z^2}
    + \frac{2 N_{z\theta}}{R}\frac{\partial^2 w}{\partial z\,\partial\theta}
    + \frac{N_{\theta\theta}}{R^2}\frac{\partial^2 w}{\partial\theta^2}.
\end{equation}

Collecting all terms on the left-hand side, the radial force balance becomes

\begin{equation}
\begin{split}
    D\nabla^4 w
    - \frac{N_{\theta\theta}}{R}
    - N_{zz}\frac{\partial^2 w}{\partial z^2}
    - \frac{2 N_{z\theta}}{R}\frac{\partial^2 w}{\partial z\,\partial\theta} \\
    - \frac{N_{\theta\theta}}{R^2}\frac{\partial^2 w}{\partial\theta^2}
    - p
    + \gamma\,\frac{\partial w}{\partial t} = 0,
\end{split}
\end{equation}
which is precisely Eq.~\ref{eq: perturbed radial displacment} when the perturbation is introduced. We assume that the axial and circumferential deformations are negligible compared to the radial displacement, and that these modes are underdamped, such that $\eta = 0$. The compatibility equation (Eq. \ref{eq: pertubed state}) can be obtained using the linearized stress-strain equations for a cylindrical shell \cite{Yamaki1984},

\begin{align}
    Eh \frac{d u}{dz}=\frac{1}{R^2}\frac{d^2 \Phi}{d \theta^2}-\nu \frac{d^2 \Phi}{dz^2} \\
    \frac{Eh}{R} \left(\frac{dv}{d \theta}-w\right)=\frac{d^2 \Phi}{dz^2}-\frac{\nu}{R^2}\frac{d^2 \Phi}{d \theta^2} \\
    Eh \left(\frac{dv}{dz}+\frac{1}{R}\frac{du}{d \theta}\right)=-2\frac{1+\nu}{R}\frac{d^2 \Phi}{d \theta dz}
\end{align}

where eliminating the axial and circumferential displacements leads to the compatibility equation,
\begin{equation}
    \nabla^4 \Phi+\frac{Eh}{R}\frac{d^2 w}{dz^2}=0.
    \label{eq: compatibiliyy eq}
\end{equation}

which is precisely Eq. \ref{eq: pertubed state} when the perturbation is introduced.

\section{Flat plate}

In this section we take the limit of a flate term where the membrane-curvature coupling terms $-\frac{1}{R}\frac{\partial^2 \Phi}{\partial z^2}$ and $\frac{Eh}{R}\frac{\partial^2 w}{\partial z^2}$ both vanish, and one recovers 
the linearised stability equations for an infinite flat plate,

\begin{align}
    D\nabla^4 w 
    - N_{xx}\frac{\partial^2 w}{\partial x^2} 
    - 2N_{xy}\frac{\partial^2 w}{\partial x\,\partial y} \nonumber \\
    \label{eq: flat plate}
    - N_{yy}\frac{\partial^2 w}{\partial y^2} +\gamma\,\frac{\partial w}{\partial t}&= 0,
\end{align}
where $x$ and $y$ are the in-plane coordinates. These are the linearised plate F\"{o}ppl--von K\'{a}rm\'{a}n equations. Studying the stability of Eq. \ref{eq: flat plate} leads to the following dimensionless growth rate,

\begin{equation}
    \sigma = -h k^4 
             + \frac{\zeta k^2}{2}\cos\!\big(2[\psi-\phi]\big).
\end{equation}
with $k^2=q_x^2+q_y^2$ and $\phi=\arctan(q_y/q_x)$. We find that any non-zero activity is sufficient to destabilize the plate, 
with the fastest-growing mode occurring at $\phi^* = \psi$ for extensile activity and $\phi^* = \psi \pm \pi/2$ for contractile activity. This demonstrates that the buckling mode orients parallel to the nematic director in the extensile case, and perpendicular to it in the contractile case.\\

\section{Simulation Details}

We simulate the nonlinear displacement equations in physical cylindrical coordinates $(z, \theta)$.  The only  non-zero component of the curvature tensor is $b_{\theta\theta} = 1/R$. 

The material parameters are fixed for all simulations with shell thickness $h = 1$, cylinder radius $R = 100$, Poisson's ratio $\nu = 0.3$, and Young's modulus $E = 1$. The nematic elastic constant is set to $L = 0.5$, nematic ordering strength is $A=1.0$, and the relaxation rates are $\eta=\gamma =  \Gamma = 1.0$ . For the extended simulations we keep the radial pressure $p = 0.0$.  The choice of these parameters do not change our results significantly. 

To numerically integrate the system, we discretize the equations on a domain of size $614 \times 2\pi R$ using a finite difference method with a grid resolution of $614 \times 614$. We apply periodic boundary conditions along the $z$-axis. Time-stepping is performed using the Euler scheme with a time step of $\Delta t = 0.05$. We run simulations for $T = 500,000  \ dt$ time steps. Further details about nematics on curved surfaces can be found in the SI. 
 
\end{document}